\documentclass{article}

 \usepackage[final]{neurips_2025_ml4ps}
\def\araa{ARA\&A}%
\def\apj{ApJ}%
\def\apjs{ApJS}%
\def\aap{A\&A}%
\def\mnras{MNRAS}%
\def\prd{Phys.~Rev.~D}%
\usepackage[utf8]{inputenc} 
\usepackage[T1]{fontenc}    
\usepackage{hyperref}       
\usepackage{url}            
\usepackage{booktabs}       
\usepackage{amsfonts}       
\usepackage{nicefrac}       
\usepackage{microtype}      
\usepackage{xcolor}         
\usepackage{graphicx}
\usepackage{subcaption}
\PassOptionsToPackage{numbers, compress}{natbib}
\usepackage[capitalize,noabbrev]{cleveref}
\title{Galactification: painting galaxies onto dark matter only simulations using a transformer-based model}

\author{%
  Shivam Pandey \\
  Department of Physics and Astronomy, Johns Hopkins University, \\
  Baltimore, MD 21218, USA. 
  \texttt{shivamp@jhu.edu} \\
  \And
  Christopher C. Lovell \\
  Kavli Institute for Cosmology, University of Cambridge, UK. \\
  \texttt{chris.lovell.astro@gmail.com} \\
  \And
  Chirag Modi \\
  Center for Cosmology and Particle Physics, New York University \\
  New York, NY 10012, USA.
  \texttt{modichirag@nyu.edu} \\
  \And
  Benjamin D. Wandelt \\
  Department of Physics and Astronomy, Johns Hopkins University, \\
  Baltimore, MD 21218, USA. 
  \texttt{wandelt@jhu.edu} \\
}

\begin{document}

\maketitle

\begin{abstract}
Connecting the formation and evolution of galaxies to the large-scale structure is crucial for interpreting cosmological observations. While hydrodynamical simulations accurately model the correlated properties of galaxies, they are computationally prohibitive to run over volumes that match modern surveys. We address this by developing a framework to rapidly generate mock galaxy catalogs conditioned on inexpensive dark-matter-only simulations. We present a multi-modal, transformer-based model that takes 3D dark matter density and velocity fields as input, and outputs a corresponding point cloud of galaxies with their physical properties. We demonstrate that our trained model faithfully reproduces a variety of galaxy summary statistics and correctly captures their variation with changes in the underlying cosmological and astrophysical parameters, making it the first accelerated forward model to capture all the relevant galaxy properties, their full spatial distribution, and their conditional dependencies in hydrosimulations. 
\end{abstract}

\section{Introduction}
\label{sec:intro}
\vspace{-0.05in}
Large-scale cosmological surveys provide statistical information on billions of galaxies, which is key to understanding the structure and evolution of the Universe. Gravitational collapse induces significant non-Gaussianity in the large-scale structure, embedding information at all orders of correlation that is difficult to capture with purely analytical frameworks. It is therefore essential to create digital analogs of these observations---in the form of simulated universes---to compare against the data. Hydrodynamical simulations, which self-consistently evolve components such as dark matter, gas, stars, and black holes, are the most physically motivated method for creating such mock galaxy catalogs. However, they are too computationally expensive to run at the scale and fidelity required to draw inferences from modern observational data.


State-of-the-art, high-resolution hydrodynamical simulations capable of resolving galaxy formation in low-mass systems are limited to box sizes of approximately 100~Mpc/$h$ \citep{Vogelsberger:2020:NatRP:, Crain:2023:ARA&A:}.\footnote{Mpc stands for mega-parsec, approximately $3.2\times 10^6$ lightyears or $2\times 10^{19}$ miles, and $h \approx 0.7$ is related to the Hubble constant.} The hydrosimulations aiming to match the cosmological observations must approximate the complex astrophysics of galaxy formation using subgrid models, whose functional forms and parameters are often poorly constrained. While larger, lower-resolution simulations are possible, a single run of these kind of simulations at a fixed set of parameters costs around $2\times 10^8$ CPU hours \citep{Pakmor:2023:MNRAS:, Crain:2023:ARA&A:}. To obtain robust constraints on our Universe's parameters within a Bayesian framework, such as through simulation-based inference \citep[SBI; see][]{Cranmer_2020}, one needs a large ensemble of simulations that span a wide range of cosmological and astrophysical models and their parameter values. This requirement exacerbates the computational challenge and strongly motivates the development of accelerated model frameworks.

\begin{figure*}[ht]
\centering
\vskip -0.15in
\includegraphics[width=0.8\textwidth, height=0.35\textwidth]{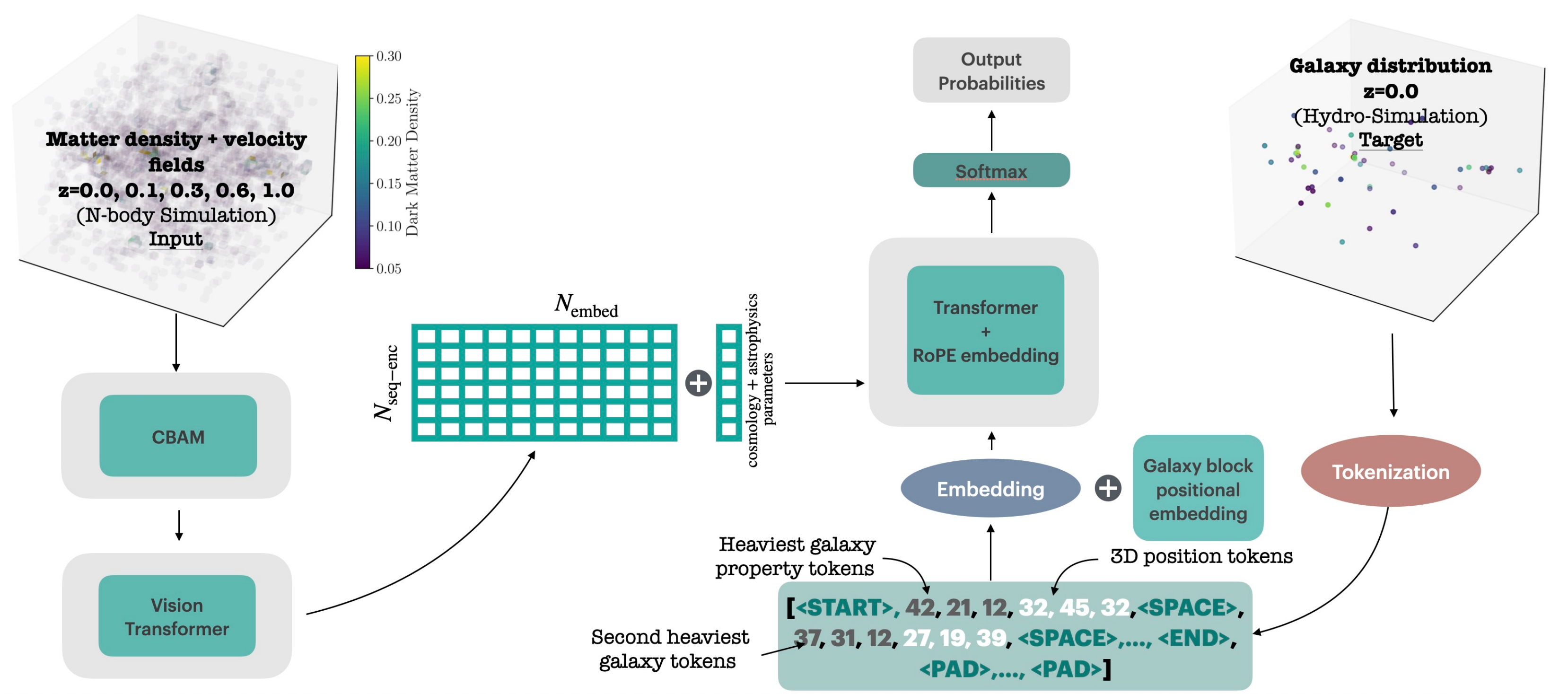} 
\caption{
\textbf{Model architecture.} Left: Input dark matter density field. Right: Target galaxy distribution. An encoder (CBAM + Vision Transformer) extracts features that condition a cross-attention decoder to generate a tokenized sequence of galaxy properties. See Sec.~\ref{sec:data_methods} and \cite{Pandey:2024:arXiv:gotham} for details.
}
\label{fig:architecture}
\end{figure*}

A much faster alternative is to simulate only the evolution of dark matter, which interacts purely through gravity. These dark-matter-only simulations, known as N-body simulations, are typically over 100 times faster than their hydrodynamical counterparts. In this work, we use N-body simulations as input to learn the distribution and properties of galaxies from a corresponding hydrodynamical simulation that shares the same initial conditions. The task is effectively to learn a high-dimensional conditional probability distribution. The transformer architecture \citep{vaswani2023attentionneed} has recently proven highly efficient for such problems, as it can interface with multi-modal inputs and outputs. We therefore adopt this approach for our framework.


\section{Related Works}
\vskip -0.05in
Machine learning methods have been widely applied to related problems in cosmological simulations and early works used deterministic mapping algorithms \citep{Zhang:2019:arXiv:, Li:2021:PNAS:, deSanti:2022:MNRAS:, Jespersen:2022:ApJ:, Chittenden:2023:MNRAS:, Hausen:2023:ApJ:}. However, as galaxy formation is an inherently stochastic process, a generative approach that captures the conditional probability distribution of galaxy properties is more appropriate.

Following this reasoning, recent studies have focused on generative models \citep{Lovell:2023:mla:, Rodrigues:2025:A&A:, Bourdin:2024:arXiv:, Cuesta-Lazaro:2024:PhRvD:, Maltz:2025:MNRAS:}. In \citet{Bourdin:2024:arXiv:}, the authors trained a score-based diffusion model to predict the distribution of galaxy counts from N-body simulations and show that a simple halo-based mapping between galaxies and N-body simulations (called the halo occupation distribution, HOD, \cite{Zheng:2005:ApJ:}) is insufficient (also see \cite{Hadzhiyska:2020:MNRAS:}). While successful for number counts, realistic mock catalogs for observational comparisons must include additional properties, such as stellar mass, velocity, and apparent magnitudes. The point-cloud diffusion model based-approaches as described in \citet{Cuesta-Lazaro:2024:PhRvD:} focused on learning the spatial distribution and properties of massive halos in N-body simulations. In principle it can be extended to learn galaxy properties but crucially they focus on a emulating a fixed number of objects for each simulation, whereas the total number of galaxies in a hydrosimulation is a strong function of the underlying cosmological and sub-grid parameters (the input parameters when running a hydrosimulation). 

In \citet{Pandey:2024:arXiv:gotham}, the authors developed a multi-modal, transformer-based model to predict the spatial distribution and properties of  \textit{halos at a fixed cosmology and redshift} in N-body simulations, conditioned on faster, approximate gravity solvers. We improve and generalize their architecture for the task of predicting galaxy distributions and properties at a fixed redshift ($z=0$) directly from N-body simulations \textit{conditioned on the varying cosmological and sub-grid parameters of the simulations}. The work presented here provides a first internally consistent way to learn the positions and properties of the galaxies (such as velocity, stellar masses and photometric magnitudes) conditioned on the parameters of the simulations while also going to smaller scales ($k \sim 10 \, h/$Mpc) compared to previous works.

\section{Data and Methodology}\label{sec:data_methods}
\vskip -0.05in
We use the Illustris-TNG Latin hypercube set from the CAMELS simulation suite \citep{Villaescusa_Navarro_2021}, which provides 1000 pairs of N-body and hydrodynamical simulations that vary two cosmological parameters (total matter density $\Omega_{\rm m}$ and matter clustering amplitude $\sigma_8$) and four astrophysical parameters governing supernova and AGN feedback.\footnote{Although here we only show results for Illustris-TNG set, we have verified that our method also works well for the Astrid set of simulations \cite{Ni:2023:ApJ:}.}\footnote{The N-body simulations are only sensitive to cosmological parameters, as they contain only dark matter and no astrophysical processes.} We divide this dataset into training (80\%), validation (12.5\%), and test (7.5\%) sets.

Our model inputs are derived from the N-body simulations. We extract dark matter density and velocity fields at five snapshots ($z = 0, 0.1, 0.3, 0.6, 1.0$) to capture the time evolution of the large-scale structure. For each simulation box of $(25~{\rm Mpc}/h)^3$, we divide the volume into eight sub-boxes and grid each field at a resolution of $16^3$. To incorporate information about the large-scale environment, we also include lower-resolution density fields from the parent box, down-sampled to match contexts of 1.5 and 3 times the sub-box size (aligned with the higher-resolution field for each sub-box). In total, 30 distinct 3D fields are concatenated along the channel dimension to form the input tensor which contains local, environment and growth of large scale structure information which is crucial to understand the galaxy formation process. 

The model's objective is to generate mock galaxy catalogs, including their 3D positions and properties: line-of-sight velocity ($v_x$), stellar mass ($\log(M_{\star})$), and SDSS g-band magnitude \citep[$M_g$; ][]{Lovell:2024:arXiv:}. We include all galaxies with stellar mass $M_{\star} > 10^{9.5} \, M_{\odot}/h$, a limit sufficient for next-generation surveys \citep{Zou:2019:ApJS:}. We scale the six properties ($x, y, z, v_x, \log(M_{\star}), M_g$) to lie in the range [0, 1] and then discretize this range into 64 bins. Each galaxy is thus represented by six tokens, forming a "word". For each sub-volume, we concatenate the tokens of all its galaxies in descending order of stellar mass to form a "sentence", which is bracketed by \texttt{START} and \texttt{END} tokens. This sequence is the target output for our model.

Our network adapts the encoder-decoder transformer architecture for cosmological simulations from \citet{Pandey:2024:arXiv:gotham} with several key modifications to learn the complicated galaxy formation process. The encoder first processes the multi-channel input fields with a Convolutional Block Attention Module \citep[CBAM;][]{woo2018cbamconvolutionalblockattention}, which uses channel and spatial attention to extract the most informative local features. The resulting feature maps are then passed to a stack of three Vision Transformer (ViT) layers \citep{dosovitskiy2021imageworth16x16words} to learn long-range correlations via self-attention. Finally, the cosmological and astrophysical parameters are appended to the ViT output, and this combined tensor is fed into the cross-attention mechanism of the decoder.

In the decoder, we first embed the galaxy tokens into a 192-dimensional space, adding a learned embedding corresponding to the token's index (1-6) to distinguish between the six different property types. We also employ Rotary Position Embeddings \citep[RoPE;][]{su2023roformerenhancedtransformerrotary} to encode the absolute position of tokens, allowing the self-attention mechanism to better capture relative dependencies. The decoder consists of 4 transformer layers with 8 attention heads. The output of the final layer is projected to predict the probability distribution for the next token in the sequence, and the model is trained by minimizing the cross-entropy loss.
\vskip -0.05in

\begin{figure*}[ht]
\centering
\includegraphics[width=0.7\textwidth, height=0.5\textwidth]{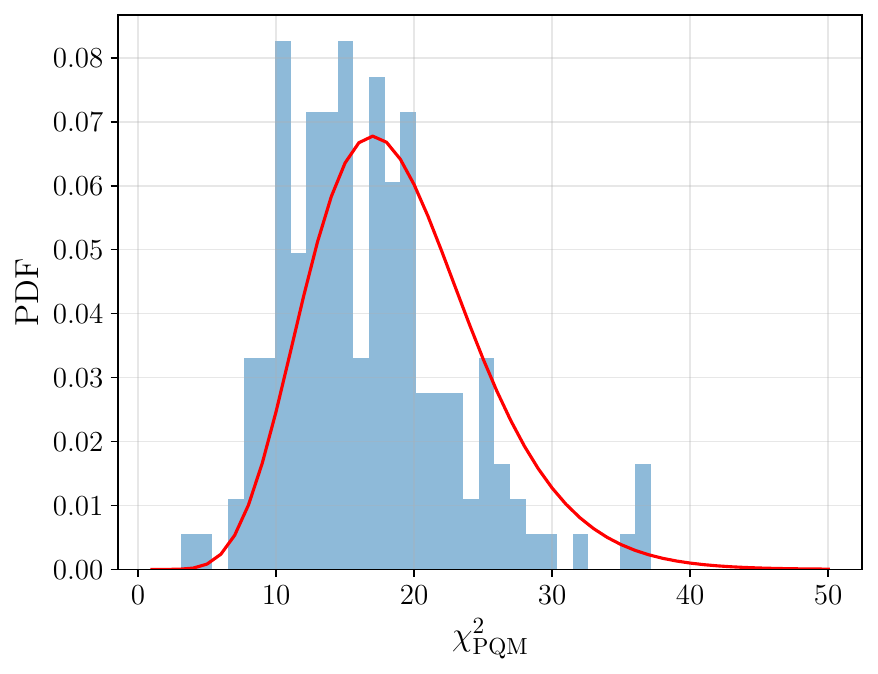} 
\caption{
\textbf{Comparison of multi-dimensional data distribution.} We each galaxy as a six dimensional vector (3 position tokens + 3 property tokens) in all the test simulations and compare the distribution of the mock and truth data using the \texttt{PQMass} methodology outlined in \cite{Lemos:2024:arXiv:}. We find that the histogram of difference between the two catalogs agrees with the red line which corresponds to the expected $\chi^2$ curve if the mock are truth come from the same underlying distribution. 
}
\label{fig:pqmass}
\end{figure*}

\begin{figure*}[ht]
\centering

\begin{subfigure}{\textwidth}
    \includegraphics[width=\textwidth]{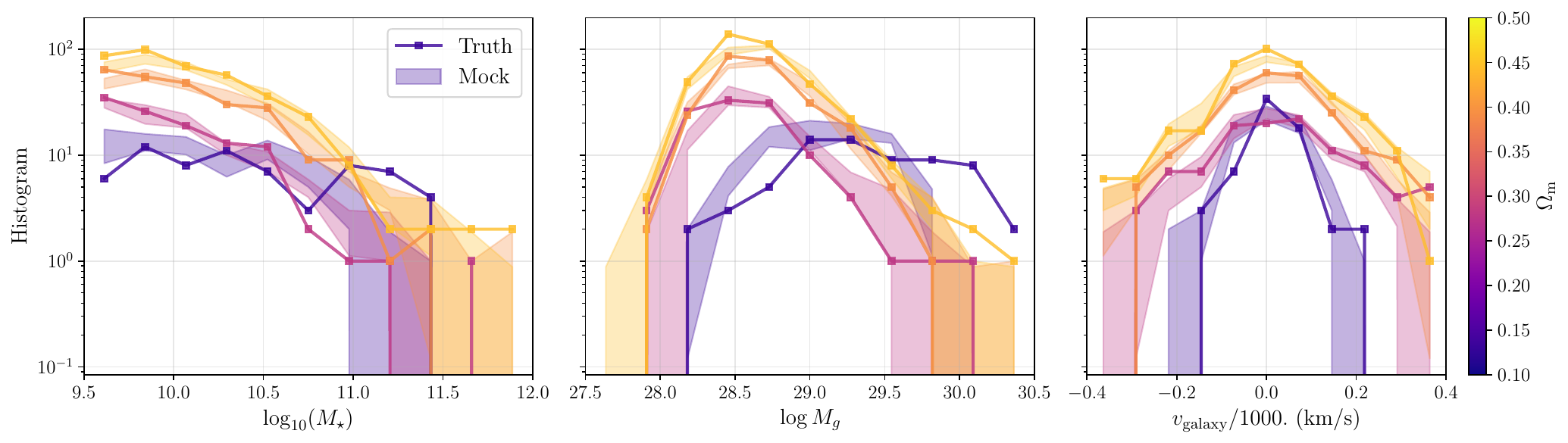}
\end{subfigure}

\begin{subfigure}{\textwidth}
    \includegraphics[width=\textwidth]{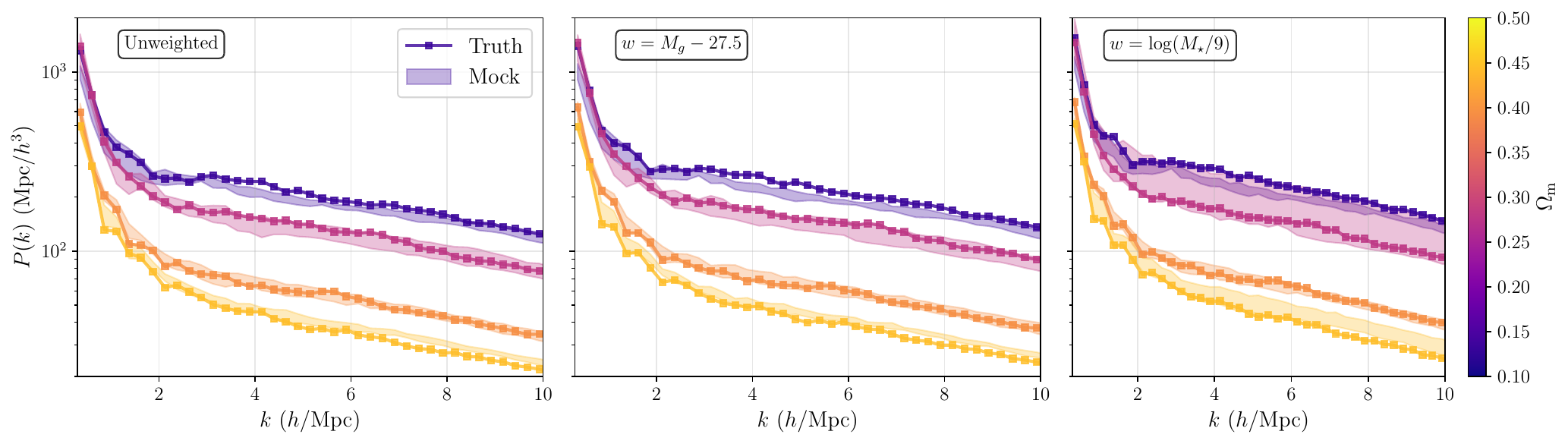}
\end{subfigure}

\caption{\textbf{Comparison of one- and two-point summary statistics.} The top row compares one-point distributions (histograms) and the bottom row compares two-point statistics. In all panels, 16th-84th percentile regions from mock catalogs sampled from our model (filled regions) are compared against the hydrodynamical simulations (truth; solid lines, squares) colored by their corresponding value of cosmological parameter $\Omega_{\rm m}$. \textbf{Top panels}: Distributions of stellar mass, g-band magnitude and line-of-sight velocity (left to right) of galaxies. Lines are colored by a cosmological parameter, showing the model captures these physical dependencies. \textbf{Bottom panels}: Redshift space power spectra, either unweighted (left), or weighted by g-band magnitude (middle) and stellar mass (right).}
\label{fig:summary_stats} 

\end{figure*}

\section{Results}
\vskip -0.05in

Once trained, we use the network to generate mock galaxy catalogs for the held-out test simulations by having the model autoregressively predict a token sequence—conditioned on dark matter fields—that is then decoded into galaxy positions and physical properties. Note that inferring the full galaxy catalog takes approximately 30 seconds on a single \texttt{Nvidia-H200} GPU, compared to 6000 CPU-hours for the equivalent hydrosimulation.

Appendix~\ref{app:mockvtrue} provides visual validations, comparing input N-body density fields with the true and predicted galaxy distributions for three different cosmologies. In Fig.~\ref{fig:pqmass} we provide a quantitative comparison between the truth and mock galaxy catalogs. We treat each galaxy as a six-dimensional vector, corresponding to 3 position coordinates and 3 properties considered in this study. Then for each held-out test simulation, we use the \texttt{PQMass} methodology described in \cite{Lemos:2024:arXiv:} to calculate the difference between the distributions of the six-dimensional data corresponding to the true and mock galaxy catalogs. The \texttt{PQMass} method partitions this sample space into non-overlapping regions and then applies $\chi^2$ tests to the number of samples residing in each region. Fig.~\ref{fig:pqmass} shows the histogram of recovered $\chi^2$ values with blue bars which agrees with the red $\chi^2$ curve (obtained for the choice of 20 regions) corresponding to the case that truth and mock data are generated from the same underlying distribution.

We first evaluate the model using one-point statistics, comparing the histograms of inferred stellar masses, g-band apparent magnitudes, and galaxy velocities against the true distributions in the top row of Fig.~\ref{fig:summary_stats}. To illustrate the model's sensitivity, the results are colored by the value of a single cosmological parameter ($\Omega_{\rm m}$), though all six parameters vary across the sample. Plotting the 16th-84th percentile region from 16 mock realizations, the figure demonstrates that our model successfully captures how changes in cosmology significantly alter these properties.

In the bottom row of Fig.~\ref{fig:summary_stats}, we probe galaxy clustering by measuring the redshift-space power spectrum. The left panel shows that the power spectra from our sampled mock catalogs (16 realizations which capture the stochasticity of galaxy formation) agree with the true spectra across various cosmologies. The middle and right panels compute power spectra weighted by g-band magnitude and stellar mass, respectively, which provide a sensitive test of the model's ability to learn the joint distribution of galaxy positions and properties, for which we find a similar level of performance. Note that we do not expect a high cross-correlation coefficient value between our mock realizations and true galaxy catalogs, especially on small scales, as our generative model captures the stochasticity of galaxy formation for a given N-body simulation.

\section{Discussion}
\vskip -0.05in

In this work, we have presented a transformer-based, multi-modal framework that generates realistic galaxy catalogs by learning the complex mapping from N-body simulations to their hydrodynamical counterparts. Our model takes dark matter density and velocity fields from inexpensive N-body simulations as input to produce a full point cloud of galaxies with associated properties (stellar mass, velocity, and magnitude), effectively acting as an accelerated forward model that reduces computational costs by a factor of ~100.

We identify several key directions for future work. A natural next step is to apply this framework to larger simulation volumes, which will increase the total number of galaxies and expand the dynamic range of their properties, providing a richer dataset for the network. We also plan to augment the model's output to include more observable properties, such as multi-band photometry and full 3D velocity vectors. To address the computational challenge of the increased context length from these enhancements, we will explore more efficient architectures, such as those incorporating sparse \citep{child2019generatinglongsequencessparse} or linear attention \citep{katharopoulos2020transformersrnnsfastautoregressive}, to accelerate both training and inference.

\section*{Acknowledgments}
This work is supported by the Simons Collaboration on ``Learning the Universe''. 
This research used the DeltaAI advanced computing and data resource, which is supported by the National Science Foundation (award OAC 2320345) and the State of Illinois. DeltaAI is a joint effort of the University of Illinois Urbana-Champaign and its National Center for Supercomputing Applications.
Some of the computations reported in this paper were also performed using resources made available by the Flatiron Institute. The Flatiron Institute is supported by the Simons Foundation. 

\bibliographystyle{mnras}
\newpage
\typeout{}
\bibliography{example_paper}
%


\appendix
\onecolumn

\begin{figure}[ht!]
    \centering
    \newlength{\totalfigheight}
    \setlength{\totalfigheight}{\dimexpr \textheight - \topskip - \headsep - \headheight - 4\baselineskip}

    \begin{subfigure}{\textwidth}
        \centering
        \includegraphics[height=0.25\totalfigheight, width=\linewidth, keepaspectratio]{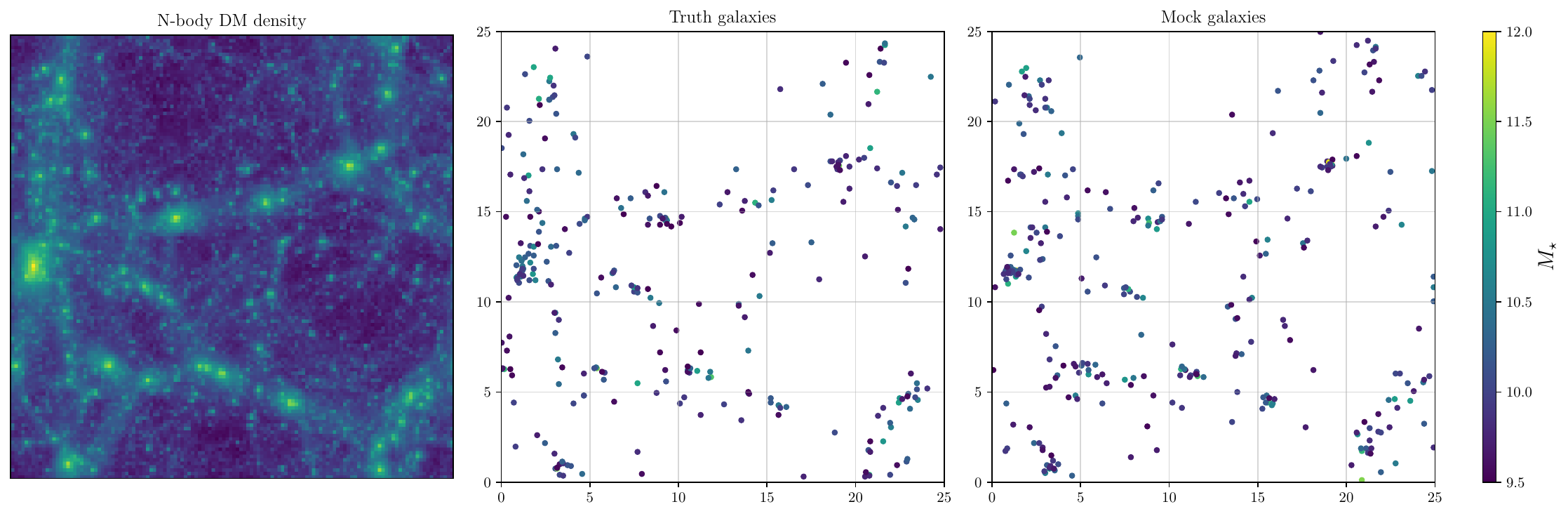}
        \label{fig:sub1}
    \end{subfigure}
    \vfill 
    \begin{subfigure}{\textwidth}
        \centering
        \includegraphics[height=0.25\totalfigheight, width=\linewidth, keepaspectratio]{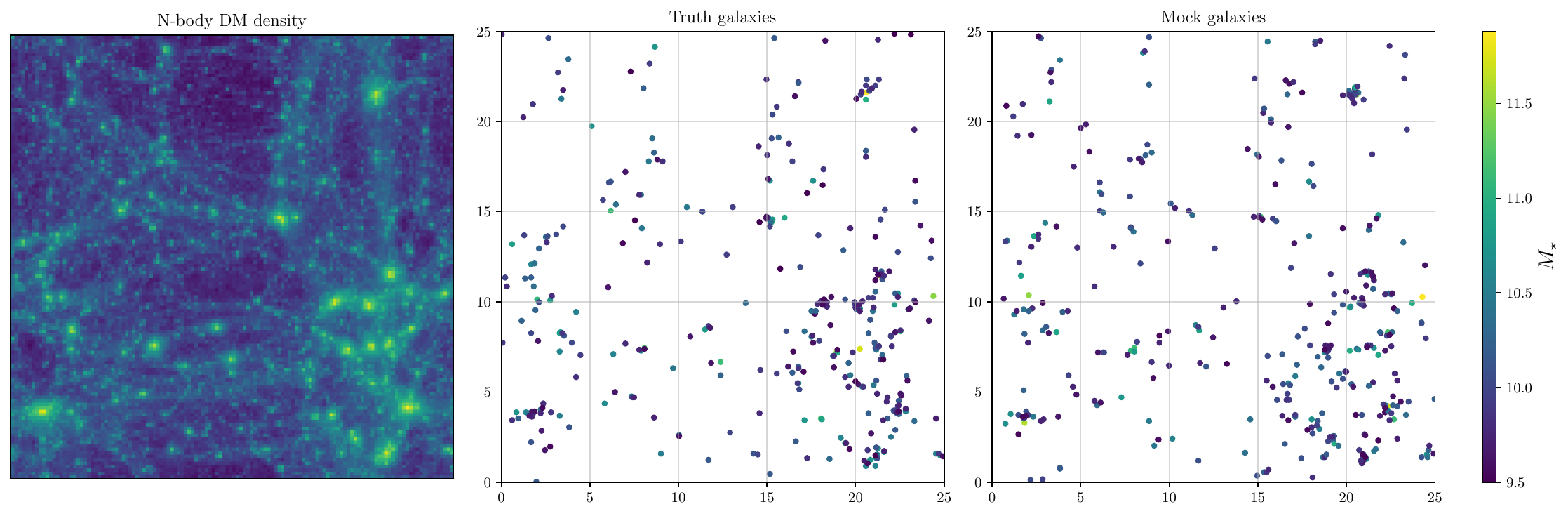}
        \label{fig:sub2}
    \end{subfigure}
    \vfill 
    \begin{subfigure}{\textwidth}
        \centering
        \includegraphics[height=0.25\totalfigheight, width=\linewidth, keepaspectratio]{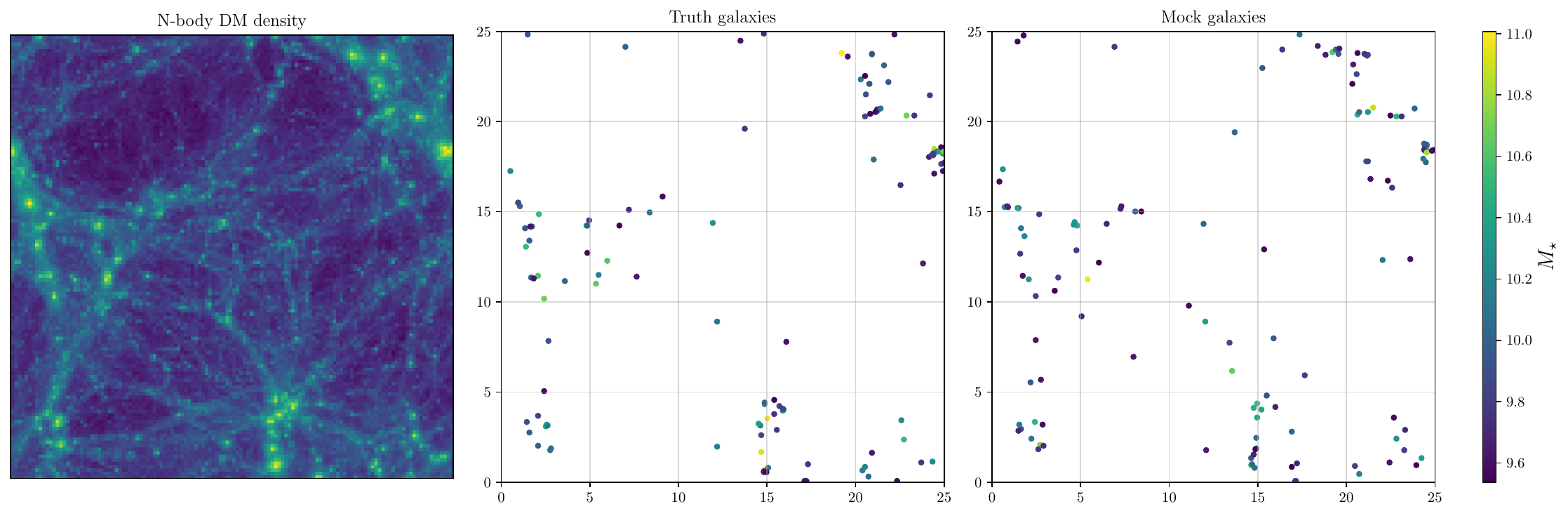}
        \label{fig:sub3}
    \end{subfigure}

    \caption{\textbf{Visual comparison of true and mock galaxy distributions.} The figure displays results for three different test simulations, each with a unique set of cosmological and astrophysical parameters (one per row). The left column shows the input dark matter density field that is one of the input fields fed to the model. The middle column shows the true galaxy distribution from the hydrodynamical simulation, while the right column shows the corresponding distribution generated by our model. In the middle and right columns, galaxies are colored by their stellar mass.}
    \label{fig:visual_comparison}
\end{figure}

\section{Visualization of the inferred catalogs}\label{app:mockvtrue}

Here we provide a qualitative validation of our model's performance with a direct visual comparison between its outputs and the ground truth. Figure~\ref{fig:visual_comparison} displays results for three distinct simulations from our test set, each with a unique combination of cosmological and astrophysical parameters. For each case (row), we show the input dark matter density field (left column), the true galaxy distribution from the hydrodynamical simulation (middle column), and the corresponding mock catalog generated by our model (right column). The galaxies are colored by their stellar mass. A visual inspection confirms that our model successfully learns to populate the dense structures of the cosmic web, generating galaxy distributions that are qualitatively indistinguishable from the ground truth across a range of underlying physical models.

\end{document}